\documentclass[aip,rsi,amsmath,amssymb,reprint]{revtex4-1}

\usepackage{graphicx}
\usepackage{dcolumn}
\usepackage{bm}

\begin{document}

\title{A parylene coating system specifically designed for producing ultra-thin films for nanoscale device applications}

\author{J.G. Gluschke}
\author{F. Richter}
\author{A.P. Micolich}
\email{adam.micolich@nanoelectronics.physics.unsw.edu.au}
\affiliation{School of Physics, University of New South Wales, Sydney NSW 2052, Australia}

\date{\today}

\begin{abstract}
We report on a parylene chemical vapor deposition system custom designed for producing ultra-thin parylene films ($5-100$~nm thickness) for use as an electrical insulator in nanoscale electronic devices, including as the gate insulator in transistors. The system features a small deposition chamber that can be isolated and purged for process termination, a quartz crystal microbalance for monitoring deposition, and a rotating angled stage to increase coating conformity. The system was mostly built from off-the-shelf vacuum fittings allowing for easy modification and reduced cost compared to commercial parylene coating systems. The production of ultra-thin parylene films for device applications is a niche not well catered to by commercial coating systems, which are typically designed to give thicker coatings (microns) with high uniformity over much larger areas. An added advantage of our design for nanoscale device applications is that the small deposition chamber is readily removable for transfer to a glovebox to enable parylene deposition onto pristine surfaces prepared in oxygen/water-free environments with minimal contamination.
\end{abstract}


\maketitle

The organic polymer, parylene, is commonly used as an encapsulant in the industrial and biomedical applications\cite{Fortin_book_2003,Ghodssi_book_2011} and has recently been deployed as a gate insulator in transistor devices featuring organic,\cite{Podzorov_APL_2003, Stassen_APL_2004, Kondo_APEX_2016} graphene,\cite{Sabri_APL_2009} transition metal dichalcogenide,\cite{Chamlagain_ACSNano_2014} and most recently, InAs nanowire\cite{Gluschke_NL_2018} conduction channels. An attractive feature of parylene compared to other organic insulators is that it is straightforwardly deposited from the gas phase using a chemical vapor deposition approach involving sublimation of a dimer precursor, high temperature cracking into reactive monomers, and room temperature polymerization exploiting drift along a pressure gradient under moderate vacuum conditions.\cite{Fortin_book_2003} However, an on-going challenge in parylene deposition is the production of ultra-thin films with good thickness control.\cite{Rapp_TSF_2012} Most commercial parylene coaters are designed for relatively thick (microns) coatings over large areas with control normally exerted solely \textit{via} the mass of starting dimer, which is sublimed to completion. Nanometer thickness films are preferred for transistor applications to reduce the gate voltage requirements, and this can entail an impractically small starting dimer mass that is difficult to accurately and reproducibly weigh out, compromising thickness control.

Previous work to obtain nanometer-thickness parylene films with good control have involved modifying commercial coating systems to incorporate thickness monitoring using either a quartz crystal microbalance\cite{Rapp_TSF_2012} or a micromachined cantilever bridge,\cite{Sutomo_JMS_2003} or a secondary chamber separated from the main chamber by an orifice to restrict monomer access.\cite{Wang_IEEE_2016,Liu_MNSL_2018} In each case, the approach is to modify a commercial system that, at delivery, was designed to do thick films on large areas with high uniformity. In contrast, our objective was to obtain ultra-thin films over small areas with excellent thickness control from the outset. The lack of an existing commercial system to modify inspired us to take a {\it de novo} approach to constructing a parylene coater tailored to our objectives. In this paper, we describe a novel design for parylene deposition systems focussed on achieving accurate thickness control of ultra-thin ($<100$~nm) parylene films for nanoscale device applications by continuously monitoring accumulated deposition thickness and designing the system to enable rapid termination of deposition once the desired thickness is achieved. This led to some substantial design differences, e.g., small deposition chamber, monomer detour circuit, and effective incorporation of previous advances, e.g., including a QCM sensor,\cite{Rapp_TSF_2012} to achieve device-functional nanometer-scale films.\cite{Gluschke_NL_2018} In this article, we describe the design and operation of our custom parylene deposition system. It was assembled from off-the-shelf parts where available, mostly using stainless steel KF and {\it Swagelok} fittings. This enables quick swapping out of components and modifications for specific projects. Our design is also economical; Our system cost approximately one third of the price of a basic commercial coater to build if we include all parts at recent market value in Australia.

The article is laid out as follows. First, we give some background on parylene coating for readers new to this material and who may be considering this as an alternative option for device insulator applications. Thereafter we describe in considerable detail the design and construction of our system. We finish with some characterization of performance and discussion of the lessons learned in designing this system.

\section{\label{sec:Background} Parylene as an electronic material}

Parylene is the trade-name for the organic polymer poly-para-xylylene. It is normally deposited from the gas phase \textit{via} a three-stage chemical vapor deposition (CVD) process involving: a) sublimation of a precursor dimer, [2,2] paracyclophane, at approximately $115^{\circ}$C, b) cracking into reactive monomers at approximately $700^{\circ}$C, and c) physisorption and polymerisation onto surfaces at room temperature.\cite{Fortin_book_2003} The process is conducted under moderate vacuum $\sim 10^{-3}$~mbar with a pressure gradient from the sublimation zone (high) to the deposition zone (low) driving drift of the gaseous dimer/monomer through the system. In addition to the standard chemistry, parylene N, a number of chemical variants exist including parylene C (mono-chlorinated), parylene D (di-chlorinated) and parylene AF-4 (fluorinated) versions, each with slightly different properties and performance advantages. All variants can be deposited by the same approach with some fine tuning of parameters.

A key feature of parylene that makes it attractive for electrical encapsulation applications is its good barrier properties, solvent resistance\cite{Koydemir_IEEE_2014, Kim_IEEE_2005}, chemical inertness\cite{Kroschwitz_book_1998, Koydemir_IEEE_2014} and high dielectric strength.\cite{Pang_IEEE_2005} Additionally, parylene is biocompatible\cite{Chang_Lang_2007} and widely used in biomedical applications such as coatings for medical implants\cite{Ghodssi_book_2011} and neural interfacing.\cite{Khodagholy_NC_2013, Koitmae_AMI_2016} Parylene is used in a wide range of industrial applications and has become an important building block of many microelectromechanical systems (MEMS).\cite{Yang_SA_1999, Yao_SA_2002, Xie_AC_2005} Parylene CVD can give exceptionally conformal polymer coatings compared to, e.g., dip coating, spin coating or spraying.\cite{Fortin_book_2003} Parylene generally adheres by physisorption, making it compatible with delicate surfaces and materials.\cite{Fortin_book_2003} This has lead to its use as an alternative gate insulator in semiconductor devices where oxides are associated with excessive charge trapping or damage to the conduction channel material.\cite{Podzorov_APL_2003, Stassen_APL_2004, Sabri_APL_2009, Chamlagain_ACSNano_2014, Gluschke_NL_2018} Thinking beyond gate insulators, ultra-thin parylene films can also be utilized as a biocompatible coating in nanoscale biosensors.\cite{Zang_CR_2016, Koitmae_AMI_2016}

Nanoscale device applications require reliable deposition of ultra-thin films, ideally with controllable thickness less than $30$~nm. Sub-$30$~nm parylene films have been previously achieved\cite{Spivack_JES_1969, Spivack_RSI_1970, Fortin_book_2003, Senkevich_CS_2003, Senkevich_CVD_2009, Rapp_TSF_2012} for film characterization but, with one recent exception,\cite{Kondo_APEX_2016} have not been used in devices. This is presumably due to two obstacles: (i) sub-$30$~nm parylene films are generally affected by pinholes,\cite{Rapp_TSF_2012} and (ii) nanometer control over film thickness is difficult to achieve with commercial coating systems. Thickness is typically controlled by varying the amount of supplied dimer; this can be inaccurate even for $>$2$~\mu$m films\cite{Sutomo_JMS_2003} due to variations in other process parameters.\cite{Kramer_JPC_1984, Fortin_book_2003, Sutomo_JMS_2003}

In a recent publication\cite{Gluschke_NL_2018} we addressed both problems. We fabricated fully functional wrap-gated nanowire field-effect transistors featuring a $20$-nm parylene~C film as the gate dielectric. The parylene film gave a conformal, electrically closed coating that could be patterned with standard nanofabrication techniques, e.g., oxygen plasma etching\cite{Meng_NT_2008} and electron-beam lithography, demonstrating the potential of parylene in nanoscale electronic device applications. Notably, our devices did not suffer from the gate leakage currents that might be expected due to pinholes, which have been previously noted for $20$~nm films with large area ($\sim 10$~mm$^{2}$).\cite{Rapp_TSF_2012} We attribute this to the sub-micron active gate area of our nanowire devices, which is much smaller than the average distance between pinholes in high-quality ultra-thin parylene films.\cite{Gluschke_NL_2018} This means that parylene can remain electrically effective in nanoscale devices even when pinhole densities increase for thinner films,\cite{Rapp_TSF_2012} providing their active area is also sufficiently small.

\begin{figure}
\includegraphics{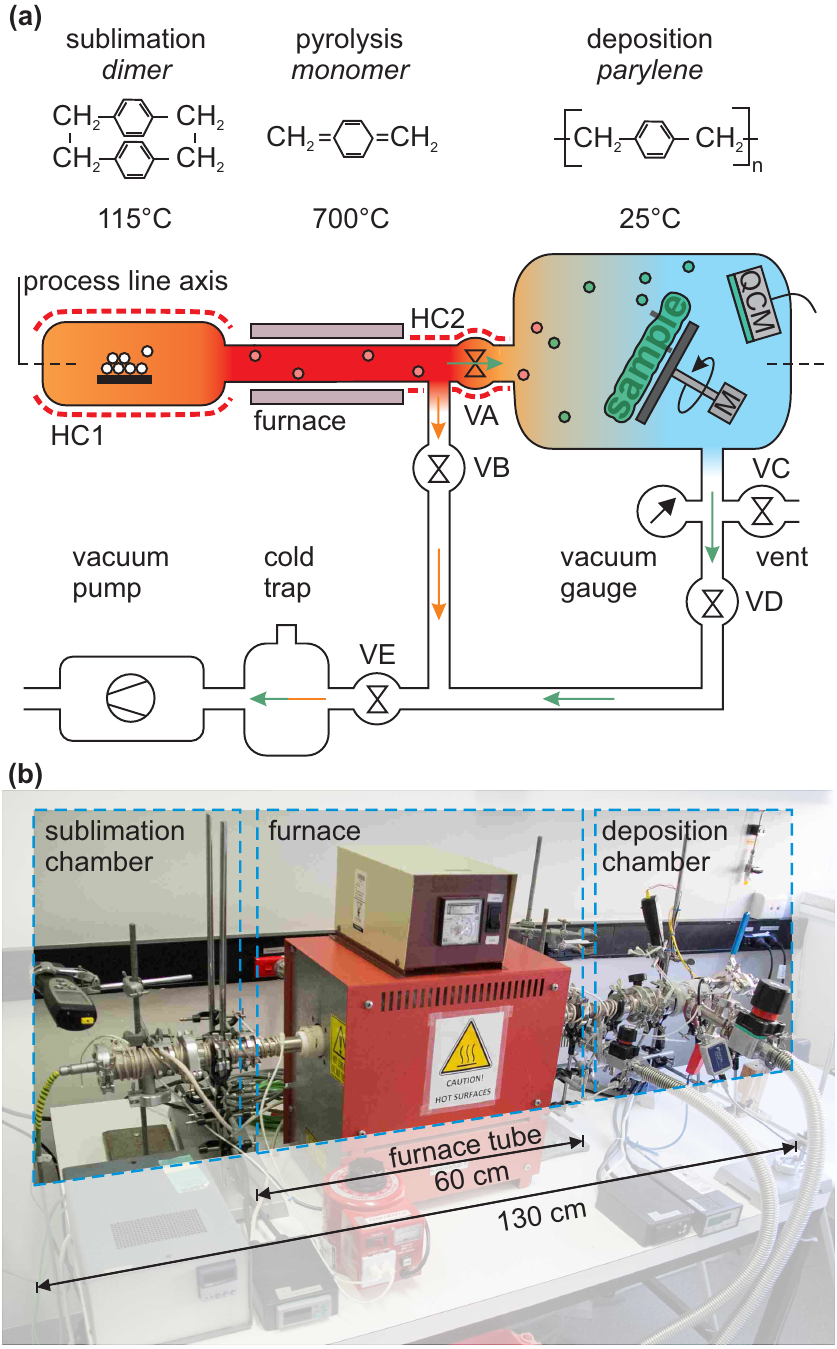}
\caption{\label{fig:F1_CVD_complete}(a) Schematic and (b) photograph illustrating the design of our parylene chemical vapor deposition (CVD) system. The system consists of three zones: sublimation, pyrolysis and deposition, moving from left to right along the process line axis. The relevant chemical species and temperatures are given above the three zones. The pumping structures are shown below, and consist of two parallel pathways to the vacuum pump. In normal operation, Valve VB and VC are closed and Valves VA, VD and VE are open, such that monomer flows \textit{via} the deposition chamber to facilitate deposition. To terminate deposition, Valve VA is closed, rapidly evacuating the deposition chamber, and thereafter Valve VB is opened to divert process flow to avoid adverse pressures in other parts of the system. The green arrows indicate the monomer flow during the deposition with Valve VA open and Valve VB closed. The orange arrows indicate the flow of monomers after closing Valve VA and opening Valve VB diverting the monomer flow around the deposition chamber. The deposition chamber also houses a quartz crystal microbalance (QCM) and motor (M) for monitoring deposition and enhancing deposition uniformity, respectively. HC1 and HC2 are heater cords heating the chamber walls to prevent unintended deposition on the walls.}
\end{figure}

\section{\label{sec:CVD} Key elements of the design of our parylene CVD system}

Figure~\ref{fig:F1_CVD_complete}(a/b) shows a schematic and photograph outlining the key design features and operating concept of our parylene CVD system. The system consists of three process zones -- sublimation, pyrolysis, and deposition -- connected upstream of a cold-trap and vacuum pump, as found in conventional parylene coating systems.\cite{Fortin_book_2003} Beyond this, our system differs in several very significant aspects. Firstly, our deposition chamber is much smaller, approximately the size of a soft-drink can ($\sim 7$~cm in diameter and $10$~cm long), with a butterfly valve (Valve VA) at the opening and a diaphragm valve (Valve VD) downstream. The small volume and butterfly valve combine to allow us to rapidly terminate deposition at any point in time -- closing Valve VA stops monomer access, allowing the pump to rapidly purge all reactive monomers from the deposition chamber. Secondly, our system has an alternate process pathway, which we activate by opening Valve VB after terminating deposition in the main deposition chamber. This diversion of monomer flow enables us to manage system pressures safely, and entirely decouple deposition control from any aspect of the sublimation process itself. This means that ultra-thin films can result from even quite large starting dimer quantities. This not only removes the inaccuracies that arise from weighing small masses of dimer, it also reduces the potential for contamination that can result if the residual impurities in the dimer sublime as the dimer runs out -- these go down a separate line to the cold-trap post-deposition, maintaining film quality. Note that it is also possible to use the alternate line at the early stages of the process to divert any volatile impurity fractions in the dimer. Thirdly, the system features a quartz crystal microbalance (QCM) to continuously monitor deposition and a small-motor drive to rotate the sample and ensure uniformity. The motor is at a $30^{\circ}$ angle to the process line axis in our system because the samples we predominantly coat are high-aspect ratio nanowires that stand perpendicular on the growth substrate, and this enables better coverage of the side-walls.

Beyond the above, the system is otherwise similar to a standard commercial parylene coater system. There is a sublimation zone with a heated boat for generating gaseous dimer, a tube furnace that serves as the pyrolysis zone, a vacuum pump with a cold trap in front of it to capture stray monomer before it can clog the pump, and an external heating system (labelled HC1 and HC2 in Fig.~1(a)) to keep the external surfaces at a temperature exceeding $100^{\circ}$C to minimise recondensation of dimer or polymerization of monomer on surfaces prior to reaching the deposition chamber.

In the sections that follow, we go into some detail regarding aspects of the design for the key sub-systems.

\section{\label{sub:Subl} Sublimation chamber}

\begin{figure}
\includegraphics{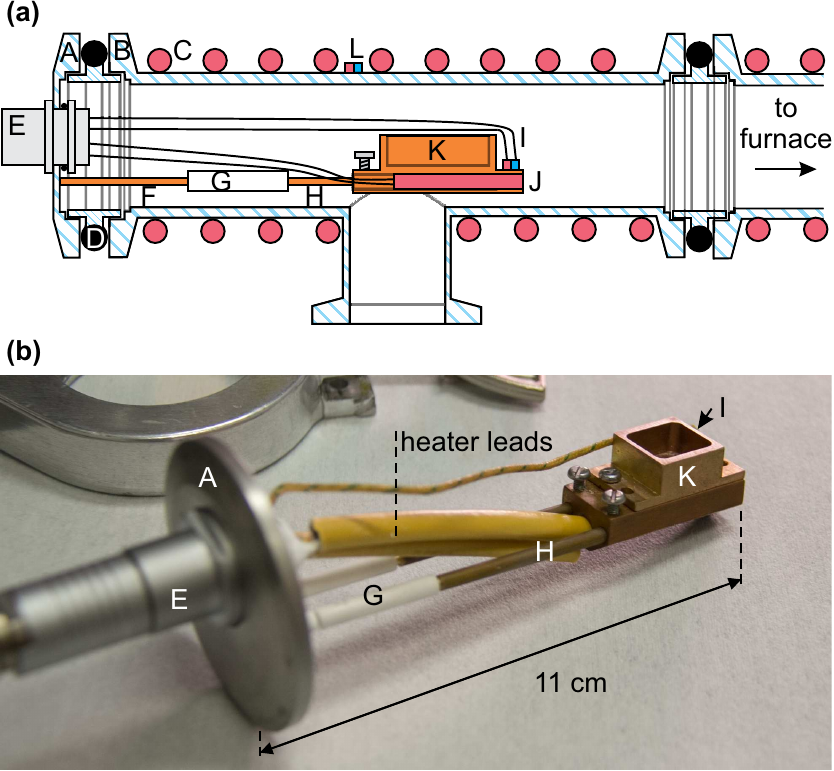}
\caption{\label{fig:F2_sublimation}(a) Schematic of the sublimation chamber; not to scale, (b) photograph of the `dimer boat' assembly. Labelled items are A: KF40 blank, B: KF40-KF16 reducing T-piece, C: {\it Briskheat} HTC451009 heating cord HC1, D: \textit{Markez} Z1028 FFKM high-temperature o-ring, E: {\it Fischer} DBEE S104-Z087-80+ connector, F: brass support rod, G: ceramic insulator, H: brass support rod, I: type-K thermocouple, J: {\it Chromalox} CIR-10121 $100~\Omega$ resistive heater, K: brass boat to carry dimer, L: Type-J thermocouple.}
\end{figure}

Figure~\ref{fig:F2_sublimation}(a) shows a schematic of the sublimation chamber. A stainless-steel KF40-KF16 reducing T-piece (B) is used as the sublimation chamber housing. The compact chamber reduces the area that needs to be heated by the heating cord HC1 (C) ({\it Briskheat} HTC451009 170~W resistive heating cord) to prevent recondensation of sublimed dimer onto the chamber walls. The external type-J thermocouple (L) is used by the HC1 heater controller ({\it Briskheat} SDC240JC-A) to maintain the external wall temperature at $>100^{\circ}$C. We use a {\it Powertech} variac SRV-5 autotransformer inserted between the HC1 controller and the heating cord to tune the heating power. The KF-40 end-plate (A) is large enough to fit a {\it Fischer} DBEE S104-Z087-80+ connector feed-through (E). This feed-through features two AWG11 solder cups with a $28$~A current rating suitable for driving the sublimation boat heater (J), which is a {\it Chromalox} CIR-10121 240~V 225~W resistive cartridge heater, $30$~mm long and $6.25$~mm in diameter. This heater fits into a cylindrical recess in the `dimer boat' (K), which was machined from a block of brass and is large enough to hold approximately $1$~g of dimer. A photograph of the dimer boat assembly is shown in Figure~\ref{fig:F2_sublimation}(b). The boat temperature is monitored using a type-K thermocouple (I) screwed onto the side of the boat; its wires pass through the two smaller solder cups in the electrical feed-through (E). This thermocouple and the sublimation heater are driven by a home-built power-supply featuring a {\it JUMO} iTRON-16 PID controller. The remaining elements are structural. The dimer boat is positioned at the center of the chamber, and mounted on a pair of three-segment support rods. These rods are brass at the two ends (F/H) joined by a tight sliding fit to a $25$~mm long ceramic segment (G) in the center. This arrangement is used to limit heat transfer between the dimer boat and chamber walls to ensure better independent control and stability of the dimer boat temperature. The additional KF16 port on the chamber housing is currently not used, but was included to provide access for a pressure gauge or additional wiring if ever required.

\section{\label{sub:Pyro} Pyrolysis furnace}

\begin{figure}
\includegraphics{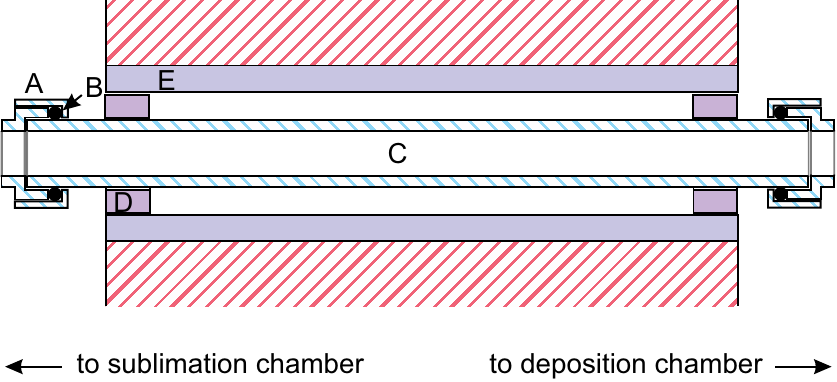}
\caption{\label{fig:F3_furnace} A: {\it Swagelok} Ultra-Torr SS-16-UT-A-20 tube fitting, B: \textit{Markez} Z1028 FFKM high temperature o-ring, C: 1-inch stainless steel tube with $1.6$~mm wall thickness, D: ceramic inserts ($27$~mm ID and $38$~mm OD), E: $40$~mm diameter furnace tube for \textit{Heraeus} RO 4/25 tube furnace.}
\end{figure}

Figure~\ref{fig:F3_furnace} shows a schematic of the furnace connected to the outlet of the sublimation chamber. This is the simplest part of the system, structurally, but nonetheless entails a significant challenge of dealing with thermal isolation between subsections in a functional way. The central component is a {\it Heraeus} RO 4/25 tube furnace (E) designed to reach temperatures up to $1200{^\circ}$C; we run this at $680-700{^\circ}$C for parylene deposition. The furnace has an integrated thermometer linked to the temperature controller. It is located on the surface of the ceramic furnace tube (external wall of ceramic cylindrical tube inside the furnace housing). A small discrepancy between the temperature at the thermometer location and the inside of the stainless-steel tube where dimer pyrolysis occurs is possible. We accounted for this discrepancy during process development by inserting a thermocouple into the stainless-steel tube and mapping the pyrolysis zone temperature against the furnace thermostat point. The pyrolysis zone temperature is typically around 5\% below the thermostat setpoint. We compensate for this in process operation by increasing the setpoint accordingly. The furnace has a ceramic tube with inner diameter $40$~mm. The process tube (C) is a $1$-inch stainless steel tube with $1.6$~mm wall thickness and length $60$~cm. Connections between the $1$~in stainless tube and the sublimation/deposition chambers are made using a pair of fittings (A) consisting of a {\it Swagelok} Ultra-Torr SS-16-UT-A-20 tube fitting welded to KF25 weld stub. The ultra-torr fittings are equipped with {\it Markez} Z1028 FFKM high-temperature o-rings (B), which are designed for use at temperatures up to $300^{\circ}$C. The process tube extends $\sim 10$~cm from the ends of the tube furnace to ensure the seals to the sublimation and deposition chambers are at an appropriately low temperature (below $200^{\circ}$C) to be well within the operating range of the {\it Markez} o-rings. The KF25 fittings connect directly to their respective chambers using standard viton o-rings. We use a pair of custom-made $5$~cm long ceramic rings (D) with inner diameter $27$~mm and outer diameter $38$~mm ({\it Mellen Co.}) to support the $1$~in stainless tube and close the gaps in the furnace tube to ensure correct operating temperature and good thermal stability in the pyrolysis zone.

\begin{figure*}
\includegraphics{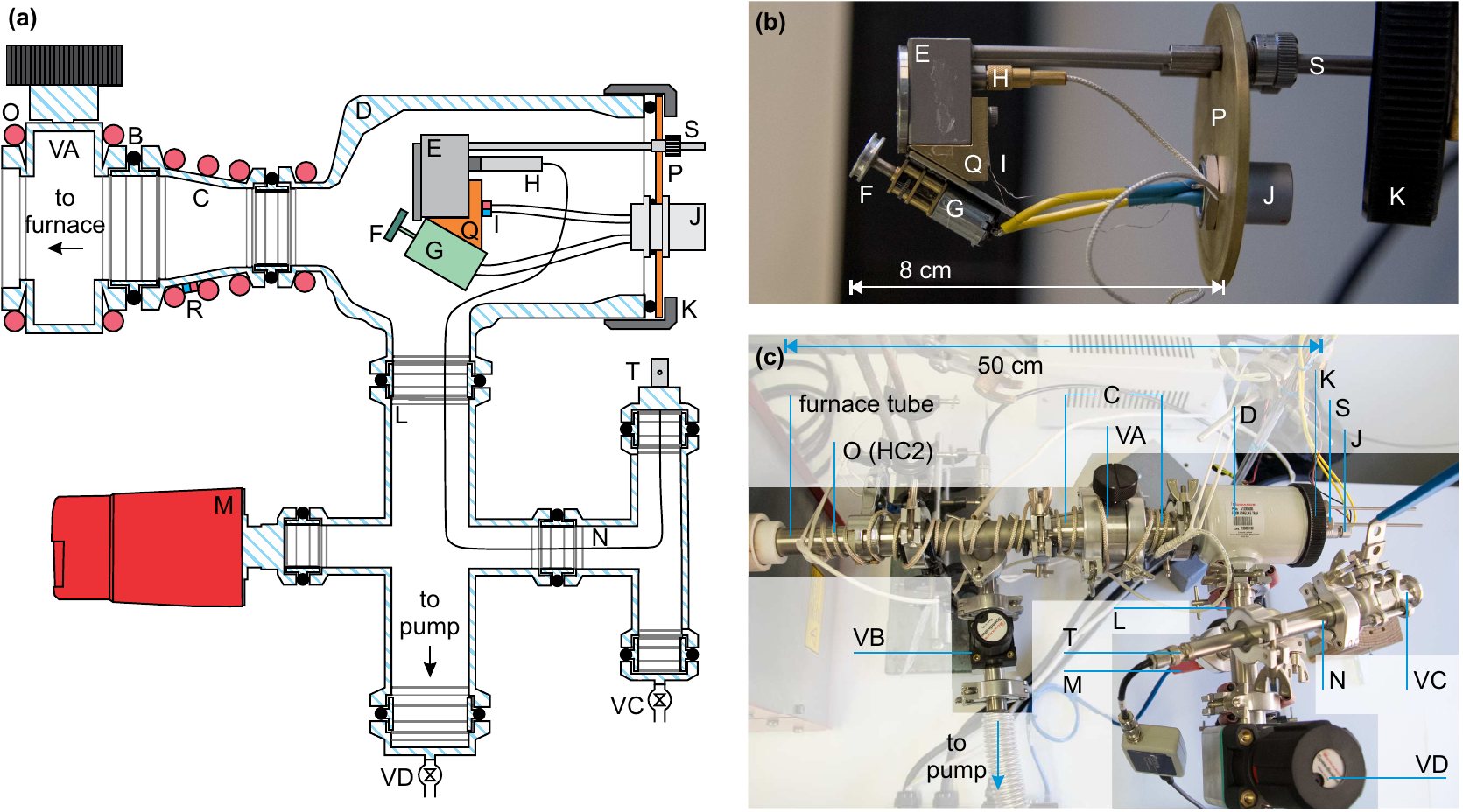}
\caption{\label{fig:F4_deposition_chamber} (a) Schematic of the deposition chamber; not to scale, (b) photograph the of sample holder and quartz crystal microbalance (QCM) sensor, (c) photograph of deposition chamber including the inlet and outlet. Labelled items are VA: {\it MDC Vacuum} 360012 KF40 butterfly valve, B: viton o-ring, C: KF40-KF25 conical reducer, D: {\it Edwards} FL20K foreline trap body, E: {\it Inficon} SL-B0E00 quartz crystal microbalance (QCM), F: SEM-stub sample holder, G: $1000:1$ low current micro metal gear-motor, H: microdot cable, I: Type-K thermocouple, J: {\it Fischer} DBEE S104-Z087-80+ electrical feed-through, K: FL20K retaining nut, L: KF25-KF16 cross-piece, M: {\it Edwards} APG100-XLC pressure sensor, N: KF16 T-piece, O: {\it Briskheat} HTC451009 heating cord HC2, P: brass end-plate, Q: brass wedge motor mount, R: Type-J thermocouple, and S: {\it Swagelok} ultra-Torr SS-2-UT-A-4 vacuum feed-throughs, T: KF16 flange MicroDot to BNC connector, VC: KF16 ball valve to vent, VD: {\it Edwards} KF25 Speedivalve for deposition chamber isolation, VB: {\it Edwards} KF16 Speedivalve.}
\end{figure*}

\section{\label{sub:Depo} Deposition chamber}

Figure~\ref{fig:F4_deposition_chamber}(a/c) shows a schematic and photograph of the deposition chamber. Figure~\ref{fig:F4_deposition_chamber}(b) shows a photograph of the sample holder assembly. The chamber housing (D) is a repurposed {\it Edwards} FL20K foreline trap, where we have replaced the aluminium top-cap with a $72$~mm diameter, $3$~mm thick brass plate (P) for feed-through access. The plate is held in place by the original retaining nut (K) and sealed by the original o-ring supplied with the FL20K. The brass plate is penetrated by two {\it Swagelok} ultra-Torr SS-2-UT-A-4 tube fittings (S) and a single {\it Fischer} DBEE S104-Z087-80+ electrical feed-through (J). The two ultra-torr fittings allow passage of the two water-cooling tubes for an {\it Inficon} front-load single sensor SL-B0E00 quartz crystal microbalance (E), which is mounted in the center of the deposition chamber. The water-cooling tubes are purely structural in this context; we do not connect them to a water supply as there is not enough heat generated by deposition for this to be necessary. A microdot cable (H), supplied with the QCM sensor, connects the sensor (E) to an external {\it Inficon} Q-pod quartz crystal monitor {\it via} a {\it McVac Manufacturing} PN 200-004 microdot-to-BNC KF16 flange feed-through (T), mounted on the Tee-piece (N) that sits below the deposition chamber. The main reason for doing this rather than going through the brass plate (P) is that a plate-mountable microdot-to-BNC feedthrough was not readily available. A pair of screw-holes on the back of the QCM sensor were used to mount a brass wedge-piece that serves two functions. First, it supports a type-K thermocouple (I) for monitoring deposition temperature. This thermocouple exits \textit{via} feed-through (J) and connects to a {\it Testo} 922 thermometer since it is not used to drive a heater. We use the remaining two pins on the feed-through (J) to drive a 1000:1 low current micro metal gear-motor (G) for sample rotation. The motor is mounted on the brass wedge piece such that its rotation axis makes a $30^{\circ}$ angle to the process-line axis, as indicated in Figure~\ref{fig:F1_CVD_complete}. The sample holder (F) is a scanning-electron microscopy (SEM) specimen stub attached to the motor drive-shaft. The motor gear ratio was chosen such that the sample rotation speed is approximately $10$~rpm, helping to ensure even conformal coating of our high-aspect ratio nanowire structures, as detailed further in Gluschke {\it et al.}\cite{Gluschke_NL_2018} The motor needs to be run at low power to avoid heating the sample, and thereby negatively impacting the deposition rate.\cite{Fortin_book_2003}

A key feature of the deposition chamber for parylene thickness control is a pair of vacuum valves: a {\it MDC Vacuum} 360012 KF40 butterfly valve (VA) in front of the deposition chamber and an {\it Edwards} KF25 Speedivalve (VD) at the back of the deposition chamber, as shown in Figure~\ref{fig:F4_deposition_chamber}(c). The KF40 butterfly valve (VA) allows rapid termination of deposition when the desired thickness is achieved without adversely restricting monomer flow into the chamber during operation. When this valve is closed, the deposition chamber is rapidly evacuated; the {\it Edwards} APG100-XLC pressure sensor (M) mounted on the cross-piece under the deposition chamber enables us to monitor this. The KF25 Speedivalve slightly restricts flow to help retain monomer supply in the chamber to drive polymerisation/deposition, and enables the chamber to be fully isolated. A final aspect of this pumping system is a KF16 ball valve (VC) that enables the deposition chamber to be vented and opened without having to break vacuum in other parts of the system (Valves VA and VD closed). This is an optional extra for convenience rather than a functional requirement.

All fittings between the furnace and deposition chamber are wrapped in Heating Cord HC2 (O) ({\it Briskheat} HTC451009 $170$~W resistive heating cord) to inhibit monomer deposition prior to reaching the deposition chamber. The type-J thermocouple (R) and heating cord are connected to a second {\it Briskheat} SDC240JC-A. We do not run the heater cord HC2 beyond the KF25 opening of the deposition chamber to avoid the hot deposition chamber walls from radiatively heating the sample, and thereby reducing the deposition rate at the sample surface. We also do not extend wall-heating downstream of the deposition chamber as any monomers that pass the chamber will obviously not return, and they are better off in the cold trap (ideal case) or pumping line walls (non-ideal case) than in the vacuum pump (worst case). This inevitably means some key components inside and downstream from the deposition chamber, e.g., the pressure sensor, valves and motor, could also accumulate thin parylene coatings over time. The pressure sensor is a Pirani gauge, which makes it immune to parylene deposition as its active components are heated during use (it measures pressure via heat loss from a hot wire). This makes it a wise choice of gauge for use in parylene coating systems in contrast to, e.g., a capacitive pressure sensor. The valves that are used unheated have relatively wide openings and readily replaceable seals. We have found after approximately 100 depositions with typical film thicknesses of 20~nm at the sample position that none of these components have been adversely affected by parylene accumulation.

\section{\label{sub:Vac} Vacuum system}

The remainder of the system from Fig.~\ref{fig:F1_CVD_complete} is relatively simple. Valves VB, VD and VE are all {\it Edwards} KF25 Speedivalves for rapid, economical operation. Valve VC is a low-cost KF16 ball valve, but could easily also be an {\it Edwards} Speedivalve or blanked off. The only critical valve in the system is Valve VA, which is a butterfly valve for reasons discussed in Section~\ref{sub:Depo}. The connections between Valves VB/VD and Valve VE are KF25 stainless steel flex-line. Valve VE is immediately in front of a liquid nitrogen cold-trap that was repurposed from a retired vacuum thermal evaporator. The cold trap is a standard feature on commercial parylene coaters also, and is designed to scavenge all remaining monomers to prevent polymerization from contaminating the vacuum pump. The vacuum pump is an {\it Edwards} RV8 rotary vacuum pump. Presently, the only vacuum gauge on the system is the {\it Edwards} APG100-XLC just behind the deposition chamber, although we have the option to add a gauge at the blanked KF16 port on the sublimation chamber (see Section~\ref{sub:lessons}), and the modular nature of the system would make it relatively easy to add pressure gauges at other points if required.

\begin{figure*}
\includegraphics{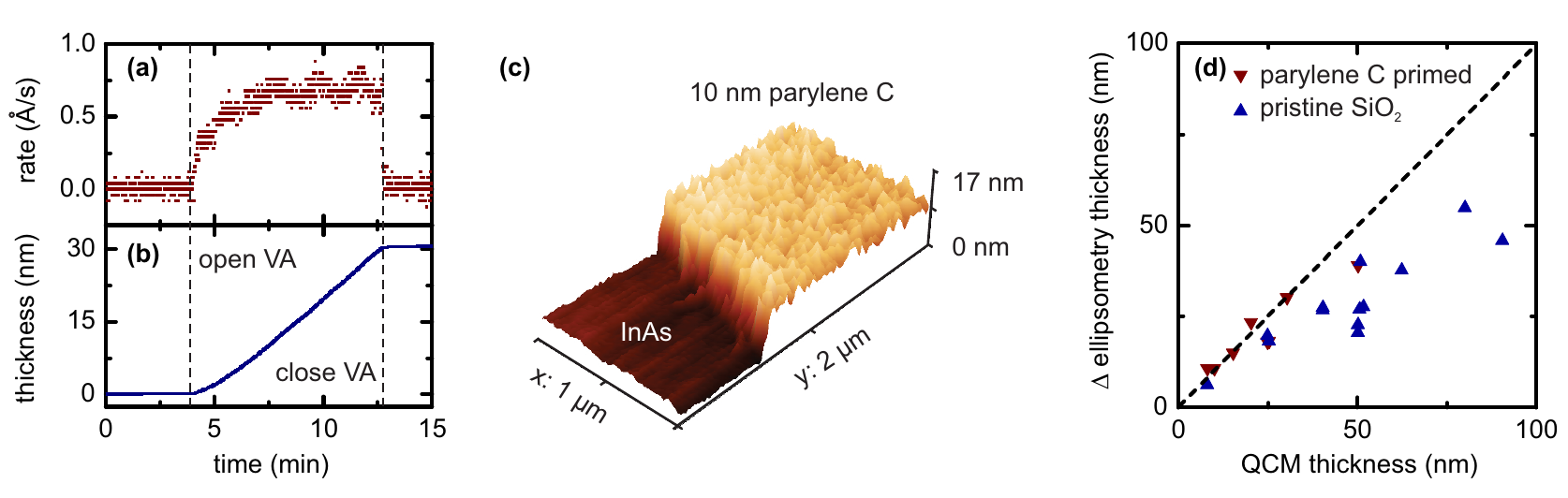}
\caption{\label{fig:F5_deposition_data} Plot of (a) parylene C deposition rate and (b) total film thickness vs time as measured by the QCM during a typical parylene deposition. The dashed vertical lines indicate the start and end of deposition upon opening and closing Valve VA. (c) AFM image of a $10$~nm thick parylene C layer on an InAs substrate. (d) Plot of the increase in film thickness measured using ellipsometry post-deposition vs thickness measured with a reused QCM crystal {\it in situ} during deposition. Pristine Si/SiO$_{2}$ substrates are shown as blue upright triangles and substrates that were pre-coated with a parylene C layer prior to this deposition are shown as red inverted triangles.}
\end{figure*}

\section{\label{sub:Rout} Typical deposition process}

The following discussion is entirely referenced to Figure~\ref{fig:F1_CVD_complete}(a) and describes a typical operating sequence. Initially, all heating elements are off and the system vented {\it via} Valve VC. The dimer boat is cleaned of any residue from previous deposition and approximately $100$~mg of {\it Curtiss Wright} parylene-C dimer or {\it SCS Coatings} DPX-N parylene-N dimer is added. The sample is mounted on the sample holder using double-sided tape. A piece of Si/SiO$_{2}$ substrate is typically mounted beside the sample for post-deposition thickness measurement using ellipsometry or AFM (see Section~\ref{sub:thickness}). The boat and sample holder are replaced in the system, which is then sealed by closing Valve VC, before evacuating to a pressure $P~=~5~\times~10^{-3}$~mbar with Valves VA and VC closed and Valves VB, VD, and VE open. The cold trap is filled with liquid nitrogen before switching on the furnace, and topped up as required. The furnace is set to 680$^{\circ}$C for parylene C and 700$^{\circ}$C for parylene N and heating cord HC2 is heated to 40$^{\circ}$C. At this point, we wait for all temperature readings to stabilize which typically takes $60-90$~min. At this point the deposition chamber pressure has typically improved to below $5~\times~10^{-4}$~mbar.

To initiate sublimation we increase the dimer boat and HC1 temperatures at approx. $\sim$10$^{\circ}$C/min to $110-120^{\circ}$C. Valve VA remains closed until the boat reaches 110$^{\circ}$C, with any dimer flow diverted along the alternate path (orange arrows in Fig.~\ref{fig:F1_CVD_complete}) so that any volatile contaminants in the dimer do not end up in the deposition chamber. Deposition is commenced by opening Valve VA and closing Valve VB. Figure~\ref{fig:F5_deposition_data}(a) and (b) shows the deposition rate and accumulated film thickness measured by the QCM during a typical deposition. Deposition starts at time $t~\approx~4$~min. The deposition rate increases to $\sim0.7$~\AA/s as the boat temperature increases from 110$^{\circ}$C to $115^{\circ}$C. Deposition stabilises at $t~\approx$~8~min as the boat temperature settles. At fixed boat temperature, we generally observe a gradual deposition rate decrease as the dimer is consumed for longer depositions than shown in Fig.~\ref{fig:F5_deposition_data}(a). The measured pressure follows roughly the same profile as in Fig.~\ref{fig:F5_deposition_data}(a) with a peak at $P~\approx~2~\times~10^{-3}$~mbar. Deposition is terminated by closing Valve VA and opening Valve VB to divert dimer flow onto the alternate path. This termination of the deposition occurs at $t~\approx~13$~min in Fig.~\ref{fig:F5_deposition_data}(a) marked by the abrupt drop in deposition rate. The heaters and furnace are all switched off and the system is left to cool under vacuum after use. Valve VC can be used to vent the deposition chamber to recover the sample without needing to vent the rest of the system with Valves VD and VA closed. Multiple sequential depositions can, in principle, be performed with this system if required.

\section{\label{sub:thickness} Thickness measurements}

The parylene film thickness after the deposition was determined by atomic force microscopy (AFM) or ellipsometry. Both methods provided reliable and consistent thickness measurements. The ellipsometer measurement proved to be more time-efficient. A $10~\times~10$~$\mu$m$^{2}$ piece of Si/SiO$_{2}$ was prepared as a reference sample by accurately determining the SiO$_{2}$ thickness using ellipsometry. The chip was placed in the deposition chamber and measured again after the deposition. The wavelengths $\lambda$ were restricted to an interval were parylene is transparent ($>$400~nm). This allows the data to be fitted as a bilayer of SiO$_{2}$ and parylene where parylene is modelled as a Cauchy material\cite{Tompkins_Ellipsometry_1993} with the refractive index $n(\lambda) = A_{n} + B_{n} \cdot \lambda ^{-2} + C_{n} \cdot \lambda ^{-4}$.\cite{Herzinger_JAP_1998} We found typical Cauchy coefficients for parylene C of $A_{n} = 1.58$, $B_{n} = 0.018$~$\mu$m$^2$ with a weak dependence on film thickness, with $C_{n}$ being negligible, in good agreement with literature \cite{Flesch_PSS_2009, Machorro_AO_1991, Callahan_JVS_2003} For thickness measurements by AFM, a reference substrate was partially masked with {\it 3M} sellotape. The tape was peeled off after the deposition leaving a sharp edge between parylene film and substrate. Figure~\ref{fig:F5_deposition_data}c shows an AFM image of a $10$~nm thick parylene film on an InAs substrate. The AFM data also shows that the film is closed with no pinholes in the imaged area. A more detailed characterization of our ultrathin parylene films and comparison to known electrical properties of thicker films is provided in Gluschke \textit{et al.}\cite{Gluschke_NL_2018} To briefly summarize, we found a breakdown electric field of 240-300~V/$\mu$m and a dielectric constant of 3.1 $\pm$ 0.7 at 120~Hz for parylene C, consistent with values obtained for bulk films.\cite{Kahouli_APL_2009,Heid_DG_2016}

An important aspect of parylene deposition is that nucleation is affected by the substrate.\cite{Vaeth_Lang_2000} Parylene readily grows on existing parylene films or surfaces treated with adhesion promoter, but can be delayed by nucleation effects on other surfaces.\cite{Vaeth_Lang_2000} The place where this has an important impact is on the QCM measurements. If the QCM sensor crystal has been used for an earlier deposition, it can accumulate parylene rapidly while deposition on the sample is delayed by nucleation, giving considerable reading errors. Figure~\ref{fig:F5_deposition_data}(d) shows the final film thickness obtained by the reused QCM vs the thickness measured by ellipsometry. We found that the QCM thickness was generally $20-30\%$ higher than actual for $\sim 30~nm$ films grown on clean untreated SiO$_{2}$ for a reused sensor crystal. Even larger discrapencies were occasionaly observed for thicker films. In contrast, for a sample with an existing parylene layer, the measured QCM thickness for a reused crystal is accurate. We recommend choosing new or used QCM crystals based on the sample, and note that crystals can be `reset' rather than replaced because parylene is readily etched in oxygen plasma. More specialised end-point sensor systems might also be useful.\cite{Sutomo_JMS_2003}

\section{\label{sub:lessons} Lessons learned}

There are several useful lessons learned in designing, building and testing this system that are useful to be aware of in either following this design or considering variations for different purposes.

Firstly, a design based on off-the-shelf vacuum fittings was a significant advantage as it can be quickly modified to improve performance or tailor it for slightly different projects. Our original intention with the small deposition chamber was to enable rapid removal of reactive monomers at the end of deposition, but we can see an additional advantage to this design. By incorporating a second butterfly or gate valve at the deposition chamber outlet, the system would enable parylene coatings onto device surfaces with high atmospheric sensitivity, e.g., graphene or transition metal dichalcogenides, without them needing to see ambient atmosphere in processing. For example, the deposition chamber can be isolated by closing both valves, removed from the system and transferred to a glovebox. A sample could be loaded under oxygen/water-free conditions, the chamber resealed, and then transferred for connection back into the parylene system. The rest of the system can be evacuated prior to opening the inlet/outlet valves for the deposition chamber, ensuring the sample surface sees only glovebox atmosphere and vacuum. This makes the design presented here potentially more useful for air-sensitive samples than commercial systems with much larger deposition chambers.

In earlier versions of our system, we used a quartz furnace tube rather than stainless steel. The strain exerted by thermal expansion and the weight of the sublimation and deposition chambers led to relatively frequent tube fracture. We find the stainless steel is far more robust with little detrimental effects from the higher thermal conductivity of stainless steel relative to quartz.

An early design concept for our system included the possibility of having a metering valve between the sublimation chamber and furnace to restrict dimer flow down the system. This is the main reason for the KF16 port on the sublimation chamber, as we intended for a second pressure gauge here to monitor sublimation chamber pressure relative to downstream deposition chamber pressure. We found this feature to be entirely unnecessary in the system as designed here, but for a system with substantive modifications, it may be necessary for constraining the parylene deposition rate to maintain adequate control for nanometer thickness parylene films.

\section{Conclusion}

We designed and built a parylene chemical vapor deposition system for controlled deposition of ultra-thin parylene films. The system features a small deposition chamber that can be isolated and purged for process termination, a quartz crystal microbalance for monitoring deposition, and a rotating angled stage to increase coating conformity. The system was mostly built from off-the-shelf vacuum fittings allowing for easy modification and reduced cost compared to commercial parylene coating systems. We recently demonstrated the capabilities of the system by fabricating functional gate-all-around nanowire field-effect transistors featuring a $20$~nm parylene C gate dielectric.\cite{Gluschke_NL_2018} The system is designed with a focus on ultra-thin films for nanoscale device applications, a niche that is not well catered to by commercial coating systems, which are traditionally designed to give thicker coatings (microns) over much larger areas with high uniformity. An added advantage of our design for nanoscale device applications is that the small deposition chamber is readily removable for transfer to a glovebox to enable parylene deposition onto pristine surfaces prepared in oxygen/water-free environments with minimal contamination.

This system was designed specifically for nanoscale devices on small $<1$~cm$^{2}$ substrates, where we can ensure that the device active area to pinhole density ratio plays in our favour. However, there are some lessons that could translate well to wafer-scale coating systems for thin parylene films. Firstly, the combination of in-situ thickness monitoring with a mechanism to rapidly isolate and purge the deposition chamber of monomer is effective in controlling the final film thickness. This would incentivise smaller deposition chambers and/or higher pumping speeds than otherwise are typical on current commercial coating systems. The key factor in scale-up with ultra-thin films will inevitably be pinholes, which are most problematic for applications requiring films with high quality over very large areas. Surface treatments prior to deposition, such as plasma etching, in combination with post-deposition annealing can help mitigate these problems.\cite{Kondo_APEX_2016} For scale-up in our application, namely nanoscale devices, pinholes become a yield issue where the yield is governed by the ratio of device active area to pinhole density. This can in principle be relatively high, as we found for our nanowire transistors, where yields of order 80\% can be obtained readily.\cite{Gluschke_NL_2018}

\begin{acknowledgments}
This work was funded by the UNSW Goldstar Scheme and the Australian Research Council (ARC) under DP170102552 and DP170104024. The work was performed in part using the NSW node of the Australian National Fabrication Facility (ANFF). We thank Jack Cochrane for contributing to designing and assembling parts of the CVD system. APM acknowledges Vitaly~Podzorov and Michael~Gershenson at Rutgers University for initial inspiration on this design. Their very simple glass tube design from ca.~2003\cite{PodzorovWeb} made us realise that building our own parylene coater rather than fighting to modify a commercial coating system was not total folly. APM and JGG also thank Colin Hall at the University of South Australia for allowing us to look inside his commercial coating system while drafting this paper to better understand the key design differences between our design and what is typically available to customers of parylene coating systems.
\end{acknowledgments}

\bibliography{Parylene_RSI}

\end{document}